\documentstyle[prb,aps,epsfig]{revtex}

\begin{document}
\draft

\title{Violation of the Fluctuation-Dissipation Theorem in a Two-Dimensional Ising Model
 with Dipolar Interactions} 

\author{Daniel A. Stariolo}

\address{Departamento de F\'{\i}sica,
Universidade Federal de Vi\c{c}osa,
36570-000 Vi\c{c}osa, MG - Brazil}
 
\author{Sergio A. Cannas\cite{auth2}}  

\address{Facultad de Matem\'atica, Astronom\'\i a y F\'\i sica, 
         Universidad Nacional de C\'ordoba, Ciudad Universitaria, 5000 
         C\'ordoba, Argentina}
\date{\today}
\maketitle

\begin{abstract}
 
The violation of the Fluctuation-Dissipation Theorem (FDT) in a two-dimensional Ising 
model with both ferromagnetic exchange 
and antiferromagnetic dipolar interactions is established and investigated 
via Monte Carlo simulations. Through the computation of the autocorrelation  
 $C(t+t_w,t_w)$ and the integrated response (susceptibility)  functions we obtain the 
$FDT$ violation factor $X(t+t_w,t_w)$ for different values of the temperature, 
the waiting 
time $t_w$ and the quotient $\delta=J_0/J_d$, $J_0$ and $J_d$ being the 
strength of exchange and dipolar interactions respectively. For positive values of 
$\delta$ this system develops a striped phase at low temperatures, in which the 
non-equilibrium dynamics  presents two different regimes. Our results show that such 
different regimes are not reflected in the $FDT$ violation factor, where $X$ goes 
always to zero for high values of $t_w$ in the aging regime, a result that appears in 
domain growth processes in non-frustrated ordered systems.

\end{abstract}


 The competition between long-range 
antiferromagnetic dipolar interactions and short-range ferromagnetic exchange 
interactions can give rise to   a variety of unusual and interesting macroscopic  
phenomena. Recent 
works in two dimensional uniaxial spin systems, where the spins are oriented 
perpendicular to the lattice and coupled with these kind of interactions, 
have shown a very rich phenomenological scenario concerning both its 
equilibrium statistical mechanics\cite{Kashuba,MacIsaac} and non-equilibrium 
dynamical properties\cite{Sampaio}. Moreover, recent results have shown some similarities
between the non-equilibrium dynamical properties of these kind of ordered 
systems and that of glassy systems\cite{Toloza}.

 Magnetization processes in these kind of 
systems are of interest due to aspects related to information storage in 
ultrathin ferromagnetic films. Moreover, there are several contexts in which 
a short-ranged tendency to order is perturbated by a long-range frustrating 
interaction. Among others, model systems of this type has been proposed to 
study avoided phase transitions in supercooled liquids\cite{Kivelson} and 
charge density waves in doped antiferromagnets\cite{Chayes,Zachar,Pryadko}

The above mentioned systems can be described by an Ising like Hamiltonian of 
the type 
\begin{equation}
\label{hamilton}
H = - \delta\sum_{<i,j>}{\sigma_i\sigma_j} + 
\sum_{(i,j)}{\frac{\sigma_i\sigma_j}{r_{ij}^3}}
\end{equation}
where the spin variable $\sigma_i=\pm 1$ is located at the site $i$ 
of a square lattice, the sum $\sum_{<i,j>}$ runs over all pairs of nearest 
neighbor sites and the sum $\sum_{(i,j)}$ runs over all distinct pair of 
sites of the lattice; $r_{ij}$ is the distance (in crystal units) between 
sites $i$ and $j$; $\delta$ represents the ratio between the exchange $J_0$ and dipolar 
$J_d$  coupling parameters, where the energy is mesured in units of $J_d$, which is 
assumed always antiferromagnetic ($J_d>0$). Hence, $\delta>0$ means ferromagnetic 
exchange copuling. 

 There are few numerical results concerning the 
equilibrium statistical mechanics, {\it i.e.} the finite temperature phase 
diagram of this model. MacIsaac and coauthors\cite{MacIsaac}
have shown that the ground state  of Hamiltonian (\ref{hamilton}) is the 
antiferromagnetic state for $\delta<0.85$. For $\delta>0.85$ the 
antiferromagnetic state becomes unstable with respect to the formation of 
striped domain structures, that is, to state configurations with spins 
aligned along a particular axis forming a ferromagnetic strip of constant 
width $h$, so that spins in adjacent  strips are anti--aligned,  forming a 
super lattice in the direction perpendicular to the strips. They also showed that striped states of increasingly higher thickness $h$ becomes more stable as $\delta$ increases from $\delta=0.85$. Moreover, they showed that the striped states are also more stable than the ferromagnetic one for arbitrary large values of $\delta$, suggesting such a phase to be the ground state of the model for $\delta>0.85$. Monte Carlo calculations on finite lattices at low temperature\cite{MacIsaac,Sampaio} gave further support to this proposal, at least for intermediate values of $\delta$. Furthermore, such simulations have shown that  striped phases of increasingly higher values of $h$ may become thermodynamically stable at {\bf finite} temperatures for intermediate values of $\delta$. This results are in agreement with other analytic ones\cite{Kashuba,Chayes}. For small values of $\delta$ 
the system presents an antiferromagnetic phase at low temperatures. At high 
temperatures, of course, the system becomes paramagnetic.

The dynamics of the model in the striped region is characterized by the formation and growth of magnetic domains, dominated by the competition between the exchange and the dipolar interactions.  Monte Carlo studies of the dynamics at low temperatures have shown the existence of two different dynamical regimes, according to the value of $\delta$. First,  for $\delta>\delta_c\sim 2.7$ the magnetization relaxes exponentially\cite{Sampaio}, with a  relaxation time that depends 
both on the temperature and $\delta$. For $\delta<\delta_c$ the magnetization 
presents a power law decay, with an exponent independent of $\delta$. 
Second, strong hysteresis effects appear\cite{Sampaio,Toloza} 
for $\delta>\delta_c$, which are almost absent for $\delta<\delta_c$. Finally, different types of aging behaviors have been observed in both regimes\cite{Toloza}.

 Aging effects, that is, history-dependence in the time evolution of correlations and 
 response functions after the system has been quenched into some non-equilibrium state, appear in a variety of ordered and disordered systems which are essentially out of equilibrium on experimental time scales\cite{Bouchaud}.
 Aging can be observed in real systems through different
experiments. A typical example is the zero-field-cooling\cite{Lundgren}
experiment, in which the sample is cooled in zero field to a sub-critical
temperature at time $t_0$. After a waiting time $t_w$ a small constant
magnetic field is applied and subsequently the time evolution of the 
magnetization is recorded. It is then observed that the longer the waiting 
time $t_w$ the slower the relaxation.

Although aging can be detected through several time-dependent quantities, a
straightforward way to establish it in a numerical simulation  is to calculate
the spin autocorrelation function
\begin{equation}
\label{autocor}
C(t+t_w,t_w) = \frac{1}{N}\sum_i\langle\sigma_i(t+t_w)\sigma_i(t_w)\rangle
\end{equation}
where $< \cdots >$ means an average over different realizations of the
thermal noise and $t_w$ is the  waiting time,  measured from some  quenching time $t_0=0$.

A second quantity of interest is the conjugated response function to an external magnetic field $h_i(t)$:
\begin{equation}
\label{reponse}
R(t+t_w,t_w) = \frac{1}{N}\sum_i\frac{\partial \langle\sigma_i(t+t_w)\rangle}{\partial h_i(t_w)}.
\end{equation}
In a variety of disordered systems and also in some domain growth processes in ordered
ones $C(t+t_w,t_w)$ and $R(t+t_w,t_w)$ are found to satisfy the generalized fluctuation-
disspation relation proposed by Cugliandolo and Kurchan\cite{Cugliandolo1}:
\begin{equation}
\label{gfdt}
R(t+t_w,t_w) = \frac{X(t+t_w,t_w)}{T}\frac{\partial C(t+t_w,t_w)}{\partial t_w }
\end{equation}

At equilibrium $C(t+t_w,t_w)$ and $R(t+t_w,t_w)$ satisfy time translational invariance 
$(TTI)$, the functions depend only on the times difference $t$ and $X(t+t_w,t_w)=1$, 
that is, Eq.(\ref{gfdt}) reduces to the usual fluctuation-disspation theorem $(FDT)$. 
Out of equilibrium such properties are not expected to hold depending on the observation time scales.
The following scenario has been proposed in the context of spin glasses
\cite{Cugliandolo1}:  for small values of $t$ ($t/t_w << 1$) the system is in quasi-equilibrium and equilibrium properties hold; in the aging regime $t/t_w >> 1$ both $TTI$ and 
$FDT$ do not hold, {\it i.e.}, $C(t+t_w,t_w)$ depends explicitly on $t$ and $t_w$ and 
$X(t+t_w,t_w)\neq 1$. Moreover, for large values of $t_w$, $X(t+t_w,t_w)$ becomes a 
function of time only through $C(t+t_w,t_w)$: $X(t+t_w,t_w)=X(C(t+t_w,t_w))$. This 
function $X(C)$ has been interpreted in terms of an effective 
temperature\cite{Cugliandolo2}.

At high temperatures $X$ equals one since the system always equilibrate at large  times 
and the equilibrium properties hold. At low temperatures, where aging phenomena appear, 
the departure of $X(C)$ from $1$ characterizes the $FDT$ violation.

This scenario has been verified in several models of spin glasses\cite{fdt_sg}, 
in the Lennard-Jones glass\cite{fdt_glass}, in kinetic Ising models\cite{sellitto} and
in polymers in random media\cite{polymers}. 
It has also been verified in the domain growth dynamics of ferromagnetic systems  
\cite{Barrat2} in dimensions $d=2$ and $3$, where $X$ has been found to be zero in the 
aging regime.

Instead of analyzing the response $R(t+t_w,t_w)$ we look at the integrated response 
function (proportional to the magnetic  susceptibility), that is, in a zero-field 
cooling numerical experiment we observe the growth of the magnetization under  a 
constant external field applied at $t_w$:

\begin{equation}
M(t+t_w,t_w) = \int_{t_w}^{t+t_w} R(t+t_w,s)\, h(s)\, ds.
\label{suscept}
\end{equation}
 
Using Eq.(\ref{gfdt}) we can rewrite Eq.(\ref{suscept}) for long times as

\begin{equation}
\frac{T}{h} M(t+t_w,t_w) = \int_{C(t+t_w,t_w)}^1 X(C)\, dC.
\label{suscept2}
\end{equation}

If $FDT$ is satisfied Eq.(\ref{suscept2}) reduces to a linear relation 
\begin{equation}
\frac{T}{h} M(t+t_w,t_w)=1-C(t+t_w,t_w),
\label{FDT}
\end{equation}
while a departure from this straight line in an $M$  {\it vs} $C$ parametric plot indicates 
a violation of $FDT$ and gives information about $X(C)$.

In this work we present the results of  Monte Carlo simulations in the
two-dimensional Ising model defined by the Hamiltonian
(\ref{hamilton}) on an $N=30\times 30$ square lattice with free boundary 
conditions. We chose the heat-bath algorithm for the spin dynamics and time 
is measured in Monte Carlo steps per site. 
For each run the system is initialized in a random 
initial configuration corresponding to a quenching from infinite temperature 
to the temperature $T$ at which the simulation is done. We compute $C(t+t_w,t_w)$ 
as a function of the observation time $t$, for 
different values of $t_w$, $\delta$ and $T$. In the striped phase at low temperatures 
this function decays quickly from $C(t_w,t_w)=1$ to a constant value that persists for 
$t/t_w<<1$; at $t > t_w$ it decays more slowly towards zero with a scaling law that 
depends on the ratio\cite{Toloza} $h(t)/h(t_w)$. The scaling function $h(t)$ appears 
to be linear for $\delta>\delta_c$ and logarithmic for $\delta<\delta_c$. 
This behavior suggests a different domain growth dynamics in every one of the dynamical 
regimes\cite{Toloza}. Hence, it is of interest to check whether such different dynamics 
are reflected in the $FDT$ violation factor or not.

At time $t_w$ we take a copy of the system, to which a random magnetic field 
$h(i)=h\, \epsilon_i$ is applied, in order to avoid favouring one the different 
phases\cite{Barrat2}; $\epsilon_i$ are taken from a bimodal distribution 
($\epsilon_i=\pm 1$) and the strenght $h$ of the field is taken small ($h=0.01$) to 
ensure linear response. We then compute the staggered magnetization\cite{Barrat2}

\begin{equation}
\label{stagger}
M(t+t_w,t_w) = \frac{1}{N}\sum_i\overline{\langle\sigma_i(t+t_w)\epsilon_i\rangle}
\end{equation}

\noindent whose conjugate field is $h$ and where the overline means an average over the 
random variables $\epsilon_i$. We then obtain parametric plots of 
$T\, M(t+t_w,t_w)/h$ {\it vs.} $C(t+t_w,t_w)$ for several values of the temperature, 
the waiting time $t_w=2^n$ ($n=5,7,9,11$) and for $\delta=2<\delta_c$ and 
$\delta=4>\delta_c$. We first made some checks at high temperatures where the system 
equilibrates quickly, verifying that both $TTI$ and $FDT$ are satisfied.

\begin{figure}
\centerline{\epsfig{file=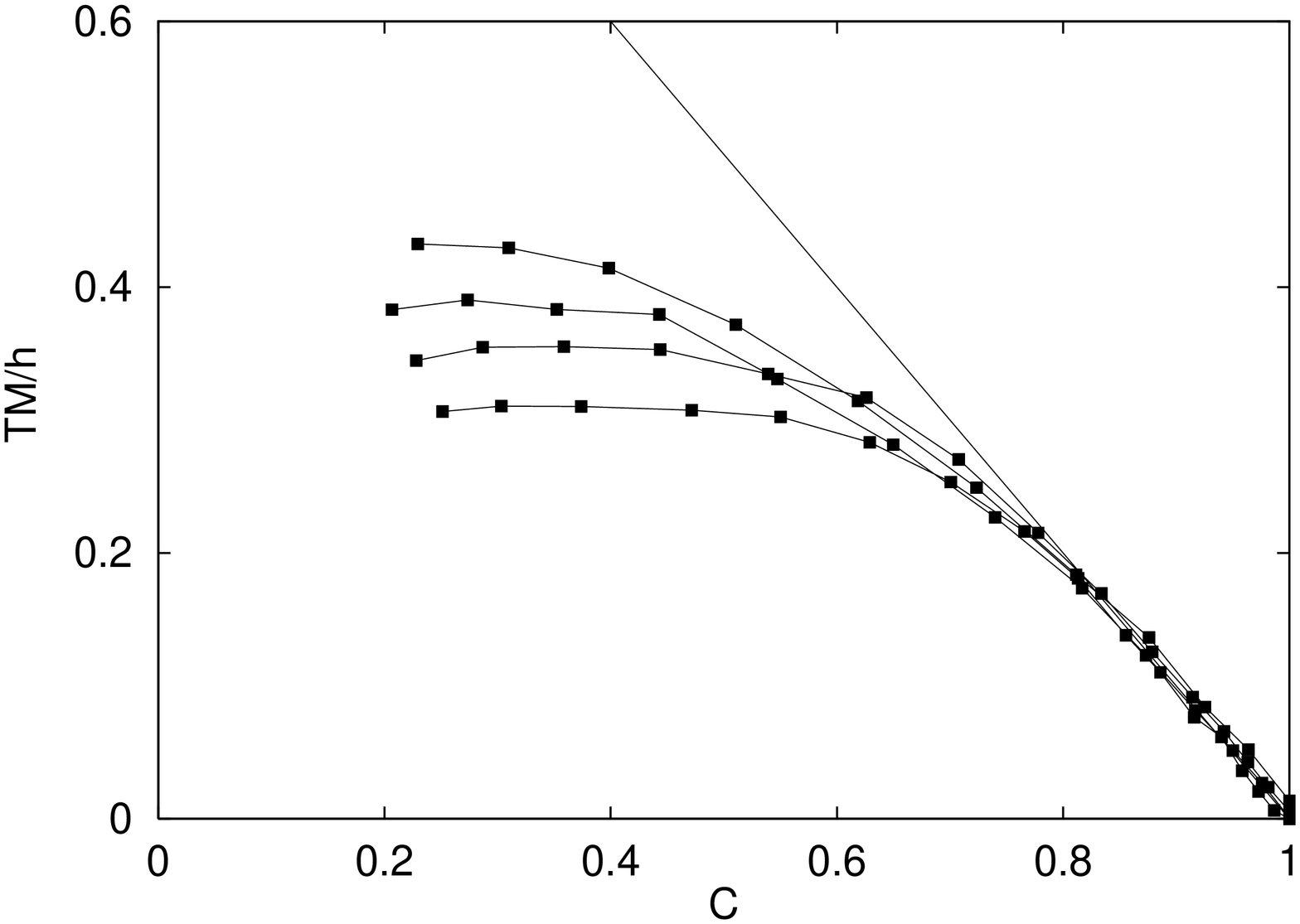,width=6cm,angle=270}}
\caption{Integrated response versus autocorrelations for $T=0.5$ and $\delta=2$.
The different curves correspond, from top to bottom, to waiting times 
$t_w=2^5, 2^7, 2^9, 2^{11}$. The straight line corresponds to the $FDT$ relation
$TM/h = 1 - C$.}
\label{fdt_d2}
\end{figure}

In Fig.(\ref{fdt_d2}) a parametric plot of the integrated response vs. autocorrelation is
shown for $\delta=2$, $T=0.5$ and waiting times $t_w=2^n$ ($n=5,7,9,11)$ from top to
bottom. We made an average over $400$ realizations of the random field. The straight
line corresponds to the $FDT$ relation (\ref{FDT}) with constant slope $-1$. For fixed
$t_w$ and observation times
$t<<t_w$ $FDT$ holds, the system is in the stationary regime. For times $t\sim t_w$ the
curves begin to depart from the $FDT$ line signalling a crossover region where the
system begins to fall out of equilibrium. Finally when $t>>t_w$ the system is out of
equilibrium with the correlations decaying to zero as $t\rightarrow \infty$. In this
last regime the integrated response keeps growing for the small $t_w$ but, as $t_w$
grows it tends to stabilize in a constant value. Furthermore, this value decreases as
$t_w$ grows. The older the system the smaller the memory of the past history. This
behavior is similar, e.g. to what happens in the coarsening dynamics of ferromagnets 
\cite{Barrat2}. It is not at all obvious that this would be the case. It is important
to note that the system is in a region $(\delta=2)$ where the ground state is the
striped phase with stripe width $h$ which grows with\cite{MacIsaac} $\delta$. For this
value of $\delta$ the scaling of the autocorrelations in the out of equilibrium regime
is logarithmic\cite{Toloza}: $C(t+t_w,t_w)\propto log(t)/log(\tau(t_w))$. 
This slow decay is typical of activated dynamics in systems with a broad distribuition
of relaxation times. This may be consequence of the degeneracy of the striped ground
state, a problem that deserves further study. So naively one would expect rather strong
memory effects as a consequence of the slow logarithmic decay in the correlations,
at variance with what is observed in the simulations. 

The flatness of the integrated response for long times implies an $FDT$ violation
factor $X(t+t_w,t_w)=0$ for $t_w\rightarrow \infty$. If we interpret $T/X$ as the
``effective temperature'' for the system in this time regime, this implies an infinite
effective temperature\cite{Cugliandolo2}.

\begin{figure}
\centerline{\epsfig{file=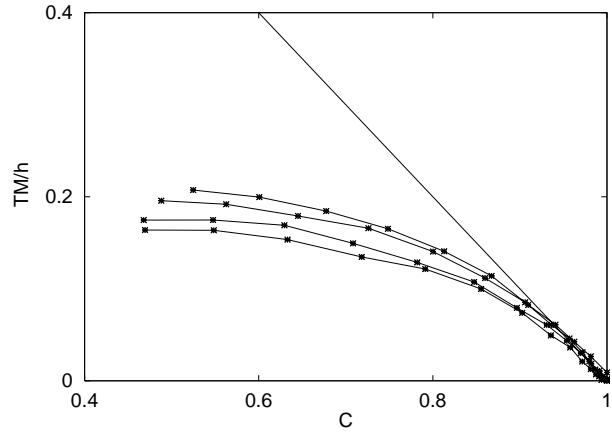,width=6cm,angle=270}}
\caption{Integrated response versus autocorrelations for $T=0.5$ and $\delta=4$.
The different curves correspond, from top to bottom, to waiting times 
$t_w=2^5, 2^7, 2^9, 2^{11}$. The straight line corresponds to the $FDT$ relation
$TM/h = 1 - C$.}
\label{fdt_d4}
\end{figure}

The $FDT$ plot for $\delta=4$ is presented in Fig.(\ref{fdt_d4}). This is 
qualitatevely similar to the plot for $\delta=2$. The main difference is that the
integrated response flattens to a value which is roughly half of that corresponding to
$\delta=2$. In this case the ferromagnetic term of the Hamiltonian is clearly
dominant. This is reflected, e.g. in the scaling form of the autocorrelations in the
aging regime\cite{Toloza}, i.e. $C(t+t_w,t_w)\propto t/\tau(t_w)$. We must note,
however, that the stable phase still corresponds to the striped one\cite{MacIsaac} but
with increasing value of the width of the stripes as $\delta$ increases. For a fixed
$t_w$ in the aging regime, the striped domains are wider with $\delta=4$ than with
$\delta=2$. Consequently, the domain walls have smaller total lenght in the latter case.
This implies a smaller contribution for the staggered magnetization while the bulk
contributions are roughly the same. So it is reasonable to expect a smaller response in
this case. In other words, as the systems becomes more ``ferromagnetic'' the long term
memory becomes weaker. 

\begin{figure}
\centerline{\epsfig{file=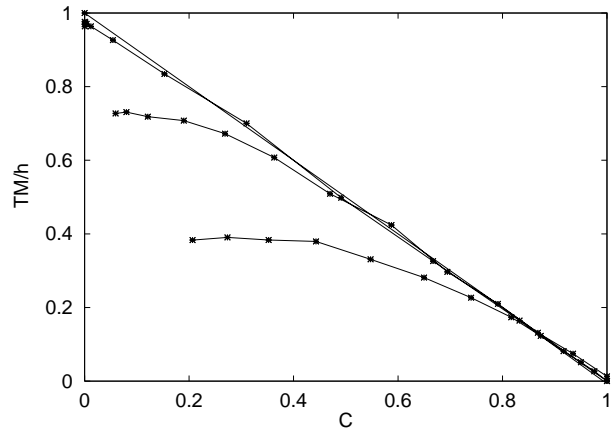,width=6cm,angle=270}}
\caption{Integrated response versus autocorrelations for $t_w=2^7$ and $\delta=2$.
The different curves correspond, from top to bottom, to temperatures
$T=1, 0.7, 0.5$. The straight line corresponds to the $FDT$ relation
$TM/h = 1 - C$.}
\label{tw7}
\end{figure}

In Fig.(\ref{tw7}) we show a parametric plot for different values of the temperature
$T=1$, $0.7$ and $0.5$ (from top to bottom) and a fixed $t_w=2^7$ for $\delta=2$.
Equilibrium dynamics is restored at $T=1$. It would be interesting to know if the change
of dynamical regime at this temperature coincides with a thermodynamic phase transition
or is a purely dynamic effect.

We have studied the $FDT$ violation in the coarsening process of an Ising model with
dipolar long range interactions. Going through $\delta_c \sim 2.7$ the aging dynamics
of the autocorrelations presents a crossover from a logarithmic decay for 
$\delta<\delta_c$ to an algebraic decay for $\delta>\delta_c$, probably related to some dynamical phase transition related with the change of the strips width. We asked weather this 
difference would manifest itself in the responses and $FDT$ violation factor $X$. It
turns out that this is not the case. For long waiting times in the aging regime 
$X\rightarrow 0$ in both cases, signalling that the long term memory is weak in both
regimes.

This work was partially supported by grants from
Consejo Nacional de Investigaciones Cient\'\i ficas y T\'ecnicas (CONICET, 
Argentina), Consejo Provincial de Investigaciones Cient\'\i ficas y 
Tecnol\'ogicas (C\'ordoba, Argentina), Secretar\'\i a de Ciencia y 
Tecnolog\'\i a de la Universidad Nacional de C\'ordoba (Argentina) and by
brazilian agencies CNPq and FAPEMIG.


\end{document}